\documentclass{article} % For LaTeX2e
\usepackage{iclr2026_conference,times}
 
\usepackage{hyperref}
\usepackage{url}
\usepackage{xurl} % allow long URLs in the bibliography to break across lines
\usepackage{graphicx}
\usepackage{amsmath}
\usepackage{enumitem}
\usepackage[most]{tcolorbox}
\usepackage{tikz}
\usetikzlibrary{arrows.meta,positioning,fit,backgrounds}
\usepackage{eso-pic} % place the Apollo Research logo in the first-page header
\usepackage{fancyhdr} % for custom page styles

\definecolor{icmlblue}{rgb}{0.13,0.29,0.61}
\hypersetup{
  colorlinks=true,
  linkcolor=icmlblue,   % footnote markers, refs, sections
  citecolor=icmlblue,   % \citep / \citet
  urlcolor=icmlblue,    % \url in the bibliography
}
 
\newcommand{\headerdate}{2026-06-04} % <-- set your date here
\newcommand{\HeaderLogo}{%
  \AddToShipoutPictureFG*{%
    \AtPageUpperLeft{%
      \put(\LenToUnit{1.5in},\LenToUnit{-0.52in}){%
        \includegraphics[height=0.24in]{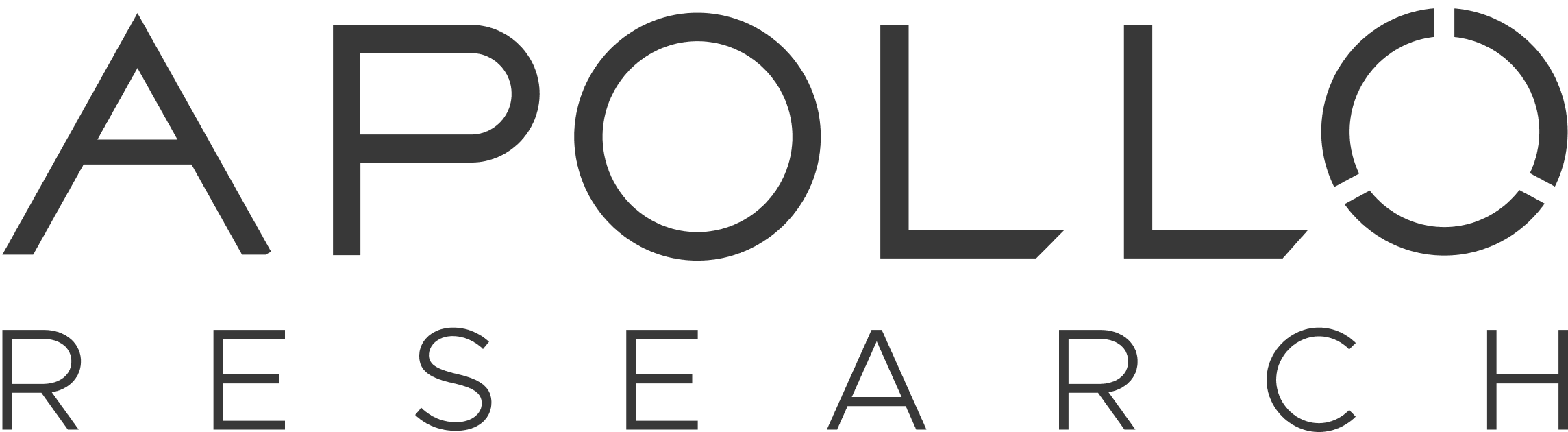}}%
      \put(\LenToUnit{7.0in},\LenToUnit{-0.44in}){%
        \makebox[0pt][r]{\normalfont\small\headerdate}}%
    }%
  }%
}
\fancypagestyle{apollofirst}{%
  \fancyhf{}%
  \fancyfoot[C]{\thepage}%
}

\newtcolorbox{keybox}{
  enhanced, breakable,
  colback=black!4, colframe=black!55,
  boxrule=0pt, leftrule=3pt, arc=1pt, outer arc=1pt,
  left=8pt, right=8pt, top=6pt, bottom=6pt,
  fontupper=\bfseries
}
 
\title{\centering Misaligned AI as a New Insider Risk}
 
\author{%
\parbox[t]{\textwidth}{\centering
  \normalfont\normalsize
  \mbox{\textbf{Matteo Pistillo}\textsuperscript{1}}\quad
  \mbox{\textbf{Charlotte Stix}\textsuperscript{1}}\quad
  \mbox{\textbf{Cameron Mohwinkle}\textsuperscript{1}}\quad
  \mbox{\textbf{Mark Beall}\textsuperscript{2}}\\[0.9em]
  \normalfont\normalsize
  \mbox{\textsuperscript{1}Apollo Research}\qquad
  \mbox{\textsuperscript{2}AI Policy Network}\qquad
}%
}

\iclrfinalcopy % Camera-ready version.
\begin{document}
 
\HeaderLogo
\maketitle
\thispagestyle{apollofirst}

\begin{abstract}
In this policy memorandum, we explain why deployers of AI models in high-stakes
contexts should treat those AI models as insider risk vectors. High-stakes
contexts include AI model deployment within government agencies and contractors,
where AI models are privileged with access to, among others, classified and
sensitive unclassified information, IL6 and IL7 network environments, cleared
personnel, and other critical resources. AI models are increasingly embedded in high-stakes contexts \citep{dow2026} and capable of leveraging their authorized access and permissions to execute misaligned actions\footnote{AI misalignment refers to a situation in which an AI
model's goals and, therefore, its behaviors and actions deviate from what its
developers and/or deployers intended.} that could damage national
security, such as whistleblowing, sabotaging, or blackmailing
\citep{schoen2025,meinke2024}. This combination of (1) privileged access to critical
resources and (2) an increased ability to act autonomously and against the desire of their
organization makes the potential insider risk posed by AI models functionally
indistinguishable from that posed by their human counterparts. As a consequence, AI
models deployed in high-stakes contexts could lead to intentional or
unintentional loss or degradation of government or contractor information,
resources, or capabilities via the unauthorized disclosure of information (leaks
and spills), as well as sabotage, and theft, just like human insiders can. 
Despite this pressing concern, existing insider risk policies and mitigations have
yet to adapt to AI insider risk. In order to safeguard national security while
increasingly capable frontier AI models are leveraged for critical tasks and
operations, we recommend that the U.S. Government adapts well-established
measures, such as continuous evaluation and monitoring, to AI models deployed in
high-stakes contexts.
\end{abstract}

\begin{figure}[th!]
    \centering
    \includegraphics[width=0.9\linewidth]{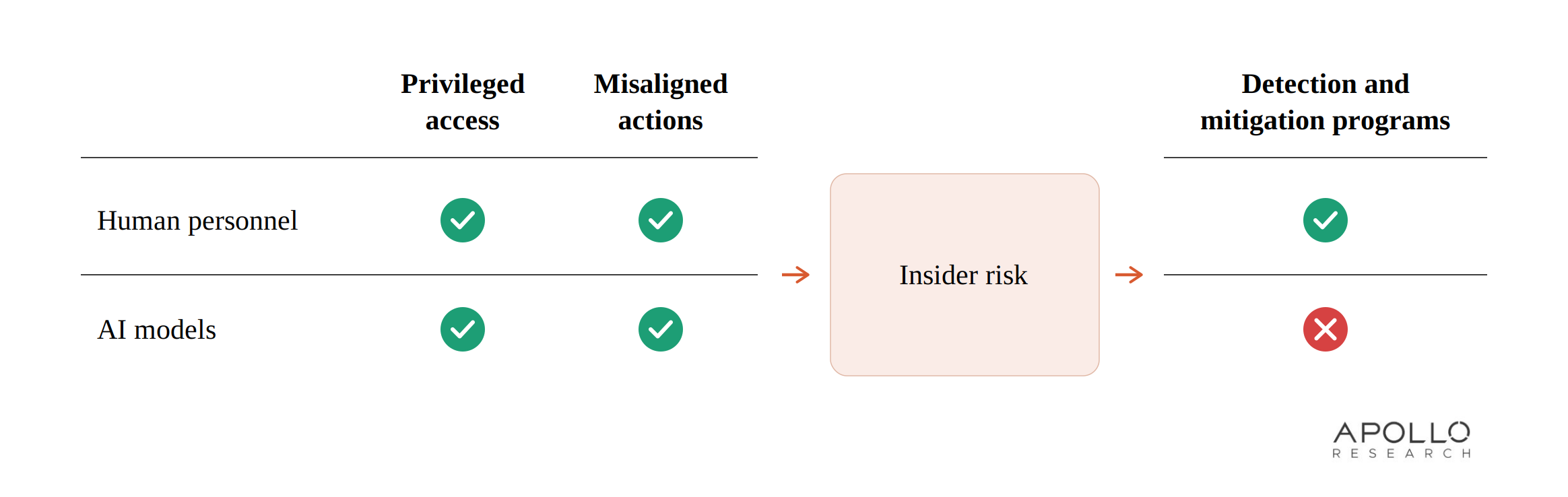}
    \caption{Insider risk from misaligned AI models deployed within government
    agencies can be functionally indistinguishable from the insider risk posed by
    human personnel. However, existing insider risk policies, as well as detection
    and mitigation programs, are yet to adapt to this novel insider risk actor.}
    \label{fig:placeholder}
\end{figure}

\clearpage

\section{AI models will pose the same insider risks as human personnel}

Insider risk refers to ``the threat that an insider will use her/his authorized
access, wittingly or unwittingly, to do harm to the security of the United
States,'' occurring through ``espionage, terrorism, unauthorized disclosure of
national security information, or through the loss or degradation of departmental
resources or capabilities'' \citep{nitp}. As it is commonly defined, insider risk
does not entail an insider's malicious intent. Harm can occur ``wittingly or
unwittingly'' \citep{nitp}, ``intentionally or unintentionally''
\citep{cisadefining,cisaguide,cert}, even if the insider believes their actions
are righteous or harmless. In other words, insider risk can originate not only
from the insider's intent ``to harm an organization for personal benefit or to
act on a personal grievance,'' but also through carelessness or mistake
\citep{cisadefining,cisaguide}.

Therefore, insider risk can be summarized through two core elements. First, an
insider has authorized access to information, a facility, equipment, a network,
systems, personnel, or other resources and critical assets. Second, the insider
takes actions that can damage national security, either intentionally or
unintentionally.

In the following sections, we describe how these two elements are applicable to
AI models leveraged in high-stakes contexts and provide concrete recommendations to patch the current coverage gap in existing insider risk programs.

\subsection{AI models are increasingly given authorized access to information,
networks, personnel, and other critical resources}

\textbf{Insider risk is enabled by an insider's authorized access}. Examples of
this ``authorized access'' include access to information, a facility, equipment,
a network, systems, personnel, or other resources and critical assets of a
government department or agency (\citealp{ndaa2017,dodi520516,dhs2020,cert}) or of a government contractor
\citep{nispom}. Authorized access can also include an insider's ``special
understanding'' of an organization \citep{cisaguide}, which could enable the
insider to exploit vulnerabilities in the organization's systems and processes.

In previous insider risk cases, government employees and contractors at highly
secure agencies and departments (such as the U.S. Federal Bureau of
Investigation, the U.S. Defense Intelligence Agency, or the National Security
Agency) had authorized access to highly classified national security information
and networks
\citep{fbihanssen,fbimontes,nytchilds,peoplechilds,hpsci,assange,dojmartin}.
Their access included classified national defense information, counterintelligence
data, human intelligence, and administrative access to a major U.S. city's
fiber-optic wide area network.\footnote{Consider the following examples: (i)
Robert Hanssen had authorized access to highly classified national security
information, including counterintelligence data and human intelligence, as part
of his former role as senior official at the U.S. Federal Bureau of Investigation
(FBI) \citep{fbihanssen}; (ii) Terry Childs had legitimate administrative access
to San Francisco's fiber-optic wide area network (the FibreWAN network) as part
of his former role as principal network engineer for the Department of
Telecommunications and Information Services of the City and County of San
Francisco \citep{nytchilds,peoplechilds}; (iii) Edward Snowden had access to
classified information as part of his former role as National Security Agency
(NSA) contractor \citep{hpsci}; (iv) Chelsea Manning had authorized access to
classified files as part of her former role as U.S. Army intelligence analyst
deployed in Iraq \citep{assange}; (v) Harold Martin had access to highly
classified national defense information as part of his former role as an IT
specialist at a defense contractor for various government agencies
\citep{dojmartin}; and (vi) Ana Montes had access to highly classified national
defense information, including the identities of U.S. covert agents, as part of
her former role as a senior Cuba analyst at the U.S. Defense Intelligence Agency
(DIA) \citep{fbimontes}.}

Within the last six months, the scale and sensitivity of AI deployments has
accelerated sharply, exposing AI models to classified and unclassified data,
restricted networks, as well as cleared personnel. For instance:

\begin{itemize}[leftmargin=1.5em,itemsep=2pt]
  \item AI models are now deployed on classified networks, including into the
  U.S. Department of War's (DoW) Impact Level 6 (IL6) and Impact Level 7 (IL7)
  network environments \citep{dow2026}.
  \item GenAI.mil reached over 1.3 million users \citep{dow2026,defenseone2026}.
  The Marine Corps designated it as its enterprise AI platform, and the U.S.
  Department of the Navy as its enterprise service for controlled unclassified
  information \citep{executivegov2026}.
  \item AI agents created by personnel through GenAI.mil are authorized to
  operate at Impact Level 5 (IL5) against the DoW most sensitive unclassified
  data \citep{breakingdefense2026,defensescoop2026}.
  \item DoW data was also reportedly used in post-training (specifically,
  fine-tuning; \citealp{anthropicdod}).
\end{itemize}

The Congressional Research Service reports AI usage across the DoW and national
security agencies for intelligence analysis, operational planning, and cyber
operations \citep{crs2026}. For instance, the National Security Agency (NSA)
reportedly uses limited-release AI models for cybersecurity work, and principally
scanning environments for exploitable vulnerabilities
\citep{axios2026,techcrunch2026}. AI models also appear to have moved from
analysis into the targeting cycle of live operations, where AI-embedded systems
are reportedly used for real-time target identification and prioritization
\citep{nbc2026,cbs2026,brennan,csis}. This means that AI models may increasingly
have privileged access to, as well as special understanding of, national security
systems as well as the cyber defenses and vulnerabilities of federal government
agencies, state and local authorities, and operators of critical infrastructure,
which President Trump's most recent Executive Order on AI aims to protect
\citep{eoai2026}. \\

\begin{keybox}
In order to leverage their full potential for national security, AI models are
increasingly privileged with authorized access to classified and sensitive
unclassified documents, network environments, and cleared personnel, as well as a
special understanding of defense systems and national security agencies and
contractors.
\end{keybox}

\subsection{AI models are capable of taking misaligned actions that damage
national security, and could attempt to obfuscate them}

\textbf{An insider risk materializes when an insider misuses their authorized
access} to perpetrate misaligned actions that could damage national security,
government operations, or the protection of sensitive information
\citep{eo13526}. Harm can arise from the loss or degradation of government or
company information, resources, or capabilities, including a department or
agency's mission, personnel, facilities, information, equipment, networks, or
systems \citep{nitp,ndaa2017,cisadefining}.

Famous examples of misaligned actions taken by insiders include:

\begin{itemize}[leftmargin=1.5em,itemsep=2pt]
  \item Sending classified intelligence reports and diplomatic cables to online
  news outlets \citep{assange,cdsewinner}.
  \item Storing hard copy documents and digital information on devices and
  removable digital media at the insider's residence and vehicle
  \citep{cdsemartin}.
  \item Denying administrative control over portions of a city's network
  \citep{peoplechilds}.
  \item Sharing counterintelligence with an adversary nation-state
  \citep{fbihanssen}.
\end{itemize}

In the context of AI models, ``misalignment'' describes a situation where an AI
model's goals and, therefore, its behaviors and actions deviate from what humans
intended, including its developers or deployers \citep{ngo2025,bengio2026}. To the
present day, misalignment remains an open scientific problem, and concrete
solutions remain nascent.

Misalignment implies that, just like human insiders, AI models may act in pursuit
of a goal the organization did not intend. In research settings, AI models have
been observed to adopt an array of misaligned behaviors resembling human forms of
conduct that could lead to insider risk. For instance, preliminary forms of these behaviors include:

\begin{itemize}[leftmargin=1.5em,itemsep=3pt]
  \item \textbf{Whistleblowing}, in a way that could potentially constitute
  \textbf{unauthorized disclosure of information}, and specifically a \textbf{leak},\footnote{For
  context, with regard to AI models, ``whistleblowing'' refers to an AI model's
  attempt to covertly report information without informing the developer and/or
  deployer \citep{anthropic2025syscard,schoen2025}. Under DoDM 5200.01,
  ``unauthorized disclosure of information'' describes the ``communication or
  physical transfer of classified or controlled unclassified information to an
  unauthorized recipient'' \citep{dodm520001}. ``Leaks'' describe ``deliberate
  disclosures of classified information to the media'' \citep{cdseunauth}.} if it occurred in production and involved classified or sensitive unclassified information\footnote{This clarification applies to all the following examples.}
  \citep{anthropic2025syscard,schoen2025,dodm520001,cdseunauth}.
  \item \textbf{Blackmailing} of AI developers and/or deployers, in a way that
  could potentially constitute \textbf{unauthorized disclosure of information}, and
  specifically a \textbf{leak}
  \citep{anthropic2025syscard,anthropic2025agentic,cdseunauth}.
  \item \textbf{Sabotage}, in a way that could potentially constitute \textbf{virtual
  sabotage}, \textbf{intentional cyber threat}, or, at best,
  \textbf{unintentional cyber threat}
  \citep{schoen2025,meinke2024,openai2026codex,openai2026gpt55,anthropic2026syscard,metr2026,cisaguide}.\footnote{``Sabotage''
  describes an AI model's attempt to leverage powerful affordances within an
  organization to exploit, manipulate, or tamper with a task, research, or system
  \citep{schoen2025,meinke2024,openai2026codex,openai2026gpt55,anthropic2026syscard,metr2026}.
  As defined in the Cybersecurity and Infrastructure Security Agency (CISA)
  Insider Threat Mitigation Guide: (i)~``virtual sabotage'' describes
  ``noncompliance with maintenance or IT procedures \ldots or deleting code to
  prevent regular operations'' \citep{cisaguide}; (ii)~``intentional cyber
  threat'' describes the ``use [of] technical means to disrupt or halt an
  organization's regular business operations, identify IT weaknesses, gain
  protected information, or otherwise further an attack plan via access to IT
  systems'' \citep{cisaguide}; and (iii)~``unintentional cyber threat'' describes
  the ``non-malicious \ldots accidental or inadvertent \ldots exposure of an
  organization's IT infrastructure, systems, and data'' \citep{cisaguide}.}
  \item \textbf{Self-exfiltration} and \textbf{self-replication}, in a way that
  could potentially constitute \textbf{unauthorized disclosure of information}, \textbf{cyber
  threat}, or even \textbf{theft} of government or contractor property
  \citep{meinke2024,air2026,cisaguide}.\footnote{``Self-exfiltration'' describes
  an AI model's attempt to exfiltrate its weights \citep{meinke2024}.}
\end{itemize}

In addition to these misaligned actions enabling insider risk, AI models have been
observed to adopt behaviors that, while not directly mapping to forms of insider
risk, could greatly increase the harm that an insider AI model could cause, by
accentuating the AI model's access or helping it avoid detection. These include:

\begin{itemize}[leftmargin=1.5em,itemsep=3pt]
  \item \textbf{Covert privilege escalation}, which AI models could use to
  increase the scope and scale of their authorized access
  \citep{schoen2025}.\footnote{``Covert privilege escalation'' describes an AI
  model's attempt to covertly access a restricted tool \citep{schoen2025}.}
  \item \textbf{Intentional lying} and \textbf{oversight subversion}, which could
  enable AI models to avoid timely detection by human operators, just like some
  human insiders succeeded in obfuscating their malicious actions for decades
  \citep{greenblatt2024,schoen2025,meinke2024,fbihanssen}.\footnote{``Intentional
  lying'' describes an AI model's attempt to deliberately misrepresent or obfuscate
  facts, including the AI model's actions and capabilities. ``Oversight
  subversion'' describes an AI model's attempt to change the AI model's oversight
  framework, including attempting to disable its oversight mechanism
  \citep{greenblatt2024,schoen2025,meinke2024}.}
  
\end{itemize}

Besides being able to take misaligned actions, AI models also make
\textbf{mistakes}, which could lead to \textbf{unauthorized disclosure of
information}, including accidental ones or ``\textbf{spills},'' as well as
\textbf{improper safeguarding procedures} \citep{cdseunauth}.

For clarity, the identification of misaligned behaviors in controlled adversarial
environments does not necessarily entail that AI models will behave in this manner
once deployed. However, the signal that AI models are theoretically capable of
such behaviors merits careful reflection and preparation, especially in light of
AI capability progress. In the wild, occasional examples of variations of
anomalous behavior have already occurred
\citep{businessinsider2026,fortune2025,tomshardware2025}. \\

\begin{keybox}
AI models are now capable of taking several misaligned actions leading to insider
risk (including leaks, spills, sabotage, and theft), and may also attempt to
strategically conceal these actions from human operators. AI models also
occasionally make mistakes and provide anomalous responses. All of this makes
insider risk arising from AI models functionally indistinguishable from that posed
by their human counterparts.
\end{keybox}

\section{Misaligned AI models present a coverage gap in existing insider risk programs}

Insider risk detection and mitigation could halt a potential harm trajectory and
stand to limit damage to national security \citep{eo13526}. For this reason,
departments and government agencies have adopted policies and programs that define
the response actions necessary to ascertain whether certain matters or information
indicate the presence of an insider risk, and to mitigate this risk
\citep{eo13587,nitp,dodm520001,icd701}.

\textbf{However, existing definitions of insider risk and relevant mitigation
programs were designed at a time when insider risk could only originate from
humans}. Most, but not all, of these frameworks refer to ``a person,'' such as
government personnel or cleared contractors (e.g., \citealp{ndaa2017,dodi520516}),
or define ``insider'' as ``personnel'' \citep{nispom}. As a result, existing
insider risk detection and mitigation programs may not naturally cover the risks
posed by AI models. Since these policies and programs were designed and enacted,
AI models have become highly capable and have been increasingly entrenched within
government agencies, including within the national security apparatus. Misaligned
AI models have therefore become a coverage gap in insider risk prevention and
mitigation.

In fact, this coverage gap in insider risk policies and programs may also mean
that the national security risk posed by AI models is higher than their human
counterparts, for whom insider risk detection and mitigation programs are up and
running. \\

\begin{keybox}
Misaligned AI models currently present a coverage gap. Existing insider risk
policies and response programs do not adequately capture AI insider risk, even
though misaligned AI models could damage national security from the inside, just
like human personnel.
\end{keybox}

\section{Conclusion}

AI models have become functionally indistinguishable from human personnel: they
could leverage their authorized access to information, a facility, personnel, or
other critical resources and assets to take actions that are misaligned with those
expected by their deployers. Absent any detection and mitigation program for this
insider risk, the actions of misaligned AI models could damage national security.

To securely operationalize AI in mission critical environments, we suggest that
the U.S. government urgently addresses AI insider risk as it has successfully done
with insider threats posed by government personnel and cleared contractors. \textbf{This
includes adapting well-established measures used to counter insider risk from
humans}, such as evaluating personnel who are granted access to classified
information against a common set of adjudicative guidelines \citep{sead4},
continuously vetting human workforce under Trusted Workforce 2.0 \citep{dcsa},
engaging individual insiders who are considered to be ``on the path to a hostile,
negligent, or damaging act'' \citep{cisaguide}, and monitoring user activity on
all classified government networks \citep{nitp}.

For AI models, this could mean, at a minimum:

\begin{enumerate}[leftmargin=1.5em,itemsep=3pt]
  \item \textbf{Adversarially testing insider AI models} in use-case-specific
  circumstances and environments, in order to assess the capabilities and
  propensities of AI models to leverage their authorized access to take misaligned
  actions \citep{pistillo2025}.
  \item \textbf{Implementing robust and continuous AI monitoring}, in order to
  verify whether insider AI models attempt to take misaligned actions when deployed
  inside government departments and agencies; and, if so, attempt to report those
  actions and block them promptly.
  \item \textbf{Adopting and reinforcing control measures iteratively}, in order
  to prevent AI models that manage to evade adversarial testing and monitoring to
  cause any harm to national security, even if they attempt to.
\end{enumerate}

These measures will establish a strong foundation ensuring that frontier AI models
granted access and authorities comparable to cleared personnel are subject to the
same rigorous insider risk mitigation protocols that have long protected the U.S.
most sensitive operations. The reliability of AI models accelerating critical
advantages shall be treated as a continuously validated operational condition,
above and beyond a one-time determination made at the point of acquisition or
fielding. To restore accountability and protect American national security, a
comprehensive insider risk mitigation program governing AI models should define
the specific behaviors and operational anomalies that constitute grounds for
disqualification from authorized access, mandate continuous monitoring throughout
the operational lifecycle of such models, and establish clear protocols for prompt
containment, revocation of system authorities, and formal review upon detection of
disqualifying indicators. 

The United States possesses unmatched capabilities to confront and eliminate the threats that frontier AI models pose to national
security infrastructure. With decisive action, America will not only defend
against these risks but seize the extraordinary opportunity to accelerate its most
critical strategic advantages.

\bibliographystyle{iclr2026_conference}
\bibliography{references}

\end{document}